\documentclass[twocolumn,showpacs,floatfix,prl]{revtex4}

\usepackage[dvips]{epsfig}

\usepackage{float}

\begin{document}

\title{Josephson Current and Noise at a Superconductor-Quantum Spin Hall Insulator-Superconductor
Junction  }

\author{Liang Fu and C.L. Kane}
\affiliation{Dept. of Physics and Astronomy, University of Pennsylvania,
Philadelphia, PA 19104}

\begin{abstract}
We study junctions between superconductors mediated by the edge states of a quantum
spin Hall insulator.  We show that such junctions exhibit a
fractional Josephson effect, in which the current phase relation has
a $4\pi$, rather than a $2\pi$ periodicity.  This effect is a
consequence of the conservation of fermion parity - the number of electrons modulo
2 - in a superconducting junction, and is closely related to the $Z_2$
topological structure of the quantum spin Hall insulator.
Inelastic processes, which violate the conservation of fermion
parity, lead to telegraph noise in the equilibrium supercurrent.
We predict that the low frequency noise due these processes diverges
exponentially with temperature $T$ as $T\rightarrow 0$.
Possible experiments on HgCdTe quantum wells will be discussed.

\end{abstract}

\maketitle

Proposals for fault tolerant topological quantum computation have motivated intense current
interest in finding robust physical systems that host excitations with non-Abelian statistics
\cite{kitaev,review}.
Recent experiments on the $\nu=5/2$ fractional quantum Hall effect have shown encouraging indirect
evidence for such excitations\cite{heiblum,marcus}, but the direct observation of non-Abelions has so far remained
elusive. Recently we showed that the proximity effect between a superconductor and a three
dimensional (3D) topological insulator leads to a 2D interface state that supports non-Abelian Majorana
fermions\cite{fukane3}.
A first step towards implementing this proposal would be to demonstrate experimentally
the topological order responsible for Majorana fermions.

In this paper we study Josephson junctions mediated by a 2D topological
insulator, also known as a quantum spin Hall insulator (QSHI)\cite{km1,km2,bernevig,bhz,konig}.
We predict that such junctions exhibit a {\it fractional Josephson effect}, whose
origin is related to the presence of Majorana fermions.  The signature of the
fractional Josephson effect is that the current phase relation
has a $4\pi$ rather than a $2\pi$ periodicity.  This behavior was first predicted by
Kitaev using an idealized model of a 1D spinless
p wave superconductor\cite{kitaev2}.  Kwon et al.\cite{kwon}
proposed that a related effect can occur at junctions between
unconventional 3D superconductors. They
argued that it leads to an AC Josephson effect with half the usual Josephson frequency and
that in a weak tunneling limit the Josephson current is carried by electrons rather
than Cooper pairs.   The $4\pi$ periodicity can occur because
the junction has two states with different Josephson currents
that are interchanged when the phase is advanced by $2\pi$.
At finite temperature inelastic processes can cause
transitions between the states, leading
 to {\it telegraph noise} in the Josephson current.  We will show that in our
setup (unlike Ref. \onlinecite{kwon}) these transitions are forbidden by the
local conservation of {\it fermion parity}, which counts the number of electrons modulo 2.
This leads to an {\it exponential}
suppression of the transition rate at low temperature.  This can be experimentally probed by
measuring the low frequency current noise $S(\omega\rightarrow 0)$, which we predict diverges
exponentially at low temperature.

The QSHI is a time reversal invariant insulating state
with a bulk energy gap generated by spin orbit
interactions\cite{km1,bhz}.  It
has recently been observed in HgCdTe quantum wells\cite{konig}.  The QSHI is
distinguished from an ordinary insulator by a $Z_2$ topological
invariant\cite{km2}, which necessitates the existence of gapless edge states.
The states at the edge of the QSHI form a unique 1D system that
is essentially {\it half} of an ordinary spin degenerate 1D electron
gas.  In the simplest case it consists of a single band of right
moving electrons paired via Kramers theorem with a left moving band
with the opposite spin.  These states are robust in the presence of
disorder because time reversal symmetry prevents elastic
backscattering.  In the absence of inelastic scattering the
transmission in the edge states is perfect.

Suppose the edge states are in intimate contact with an s wave
superconductor.  Due to the proximity effect, the tunneling of Cooper
pairs will induce a pairing potential $\Delta = \Delta_0 e^{i\phi}$ in the edge
states, which depends on the phase $\phi$ of the superconductor and
the nature of the contact.  Using the notation of Ref. \onlinecite{fukane3} write
$H = \Psi^\dagger {\cal H}\Psi/2$, where $\Psi =
[(\psi_\uparrow,\psi_\downarrow),(\psi^\dagger_\downarrow,-\psi^\dagger_\uparrow)]$
is express in terms of field operators $\psi_{\uparrow(\downarrow)}$
describing the right(left) movers and
\begin{equation}
{\cal H} =  -i v\tau_z\sigma_z \partial_x - \mu\tau_z + \Delta_0(
\cos\phi \tau_x + \sin\phi \tau_y).
\label{h0}
\end{equation}
$\sigma_j$ are Pauli matrices acting in the space of right and left movers
$\psi_{\uparrow,\downarrow}$ and $\tau_j$ are Pauli matrices which mix the $\psi$ and
$\psi^\dagger$ blocks of $\Psi$. $v$ is the velocity of the edge states, $\mu$ is the
chemical potential and we set $\hbar=1$.
The eigenstates
of (\ref{h0}) come in pairs at $\pm E$.  Due to the redundancy in $\Psi$, these states are
not independent, and the Bogoliubov quasiparticle operators satisfy $\Gamma_{-E} =
\Gamma_E^\dagger$.

Eq. \ref{h0} is similar to Kitaev's model of superconducting
spinless electrons in 1D\cite{kitaev2}.  In Kitaev's model there are
zero energy Majorana bound states associated with the {\it ends} of
the sample.  In our system, the edge - which is the boundary of the 2D
QSHI - can not have an end.  By breaking time reversal symmetry,
however, a Zeeman field can introduce a mass term into $H$ of the
form
\begin{equation}
V_Z = M \psi^\dagger \sigma_x \psi = M\Psi^\dagger \sigma_x
\Psi/2.
\label{vz}
\end{equation}
When $M>\mu$, $V_Z$ opens an insulating gap in the edge state
spectrum.  $V_Z$ could arise either
from an applied magnetic field (as in Ref. \cite{konig}) or
due to proximity to a magnetic material.  Zero energy Majorana bound
states will exist at the interface between regions with gaps
dominated by $\Delta$ and $M$\cite{fukane3}.  In the presence of both
$\Delta$ and $M$ the gap is the smaller of $|\Delta_0 \pm M|$.
When $\Delta_0 = |M|$ a single band is gapless, and for
$\Delta_0 \sim |M|$ the low energy sector of (\ref{h0}) has the form of a Su-Schrieffer-Heeger
model\cite{ssh}, which has a well known zero energy bound state where
$\Delta_0 - |M|$ changes sign.  The Bogoliubov quasiparticle operator
associated with this state is a Majorana fermion, which
satisfies $\gamma_0 = \gamma_0^\dagger$.

\begin{figure}
\centerline{ \epsfig{figure=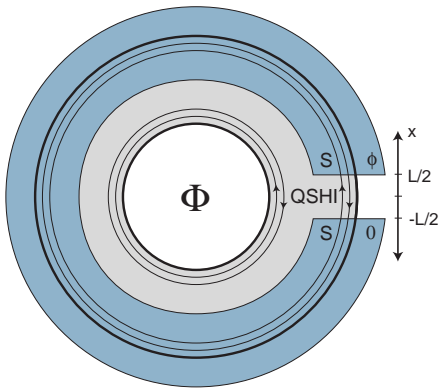,height=2in} }
 \caption{A S-QSHI-S junction in an RF SQUID geometry where the
 QSHI forms a Corbino disk. }
 \label{corbino}
\end{figure}

Consider a superconductor-QSHI-superconductor (S-QSHI-S) junction in which
the edge states of a QSHI connect two superconductors separated by a
distance $L$.  Fig. 1 shows
an RF SQUID geometry, in which the phase difference across the
junction $\phi=(2e/\hbar)\Phi$ is controlled by the magnetic flux $\Phi$.
We also assume that the QSHI forms a Corbino disk
which circles the flux.
As a practical matter, this geometry is not
essential, but we will see that it has considerable
conceptual value.  We will also consider the effect of a Zeeman term in the gap between
the superconductors, which will make the connection with Majorana bound states
transparent.  We emphasize, however, that there will be a non trivial effect even when this term
is absent.  To determine the characteristics of the junction we
solve the Bogoliubov de Gennes (BdG) equation with
\begin{eqnarray}
\Delta(x) &=& \Delta_0 \left[\theta(-x-L/2) + e^{i\phi} \theta(x-L/2)\right] \nonumber \\
M(x) &=& M_0 \theta(x+L/2) \theta(-x+L/2).
\label{dm}
\end{eqnarray}
By matching the
solutions it is straightforward to determine the spectrum of Andreev bound states
in the junction.  The calculation is similar to
Ref. \onlinecite{kwon}, as well as the theory of superconducting quantum point
contacts(SQPCs)
\cite{review2,beenakker}.  However, we shall see
that there is a fundamental difference with both of those theories.

\begin{figure}
\centerline{ \epsfig{figure=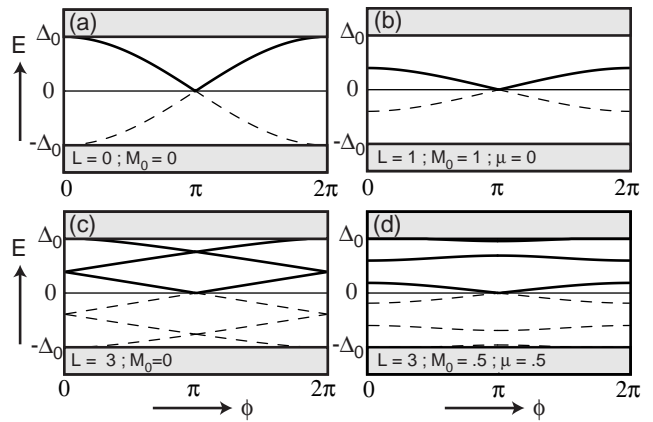,width=3.3in} }
 \caption{Spectrum of Andreev bound states in the junction as a function of
 phase difference $\phi$ for parameters indicated in each
 panel.  $L$ is in units of $v/\Delta_0$ and $M_0$ and $\mu$ are in units of $\Delta_0$.
 (a) and (c) are independent of $\mu$.}
 \label{spectrum}
\end{figure}

Fig. 2a shows the spectrum as a function of $\phi$ for $M_0=0$.  For $L \lesssim v/\Delta_0$
there is a single pair of bound states $E = \pm\epsilon_0(\phi)$.
For $L\ll v/\Delta_0$ our model reduces to the $\delta$ function model solved in Ref.
\onlinecite{kwon}, where the normal state transmission probability is
$D = 1/[1+(M_0\sinh(\kappa L)/\kappa)^2]$,
with $\kappa = \sqrt{M_0^2-\mu^2}$.  In that case
\begin{equation}
\epsilon_0(\phi) = \sqrt{D}\Delta_0 \cos(\phi/2).
\label{epsilon0}
\end{equation}
Fig. 2b shows a case where $M_0 \sim \Delta_0$, so the normal state transmission $D<1$.
When $D\ll 1$ there are two weakly coupled Majorana end states at $x=\pm L/2$.
When $L > v/\Delta_0$ there will be additional Andreev bound states in the junction with a level
spacing of order $v/L$.  Fig. 2c shows the case where $vL/\Delta_0=3$ with $M_0=0$, in which
time reversal symmetry requires Kramers degeneracies when $\phi = 0$ or $\pi$.  Fig. 2d shows the
effect of finite $M_0$ and $\mu$, which lifts most of the degeneracies.  However, the crossing
at $E=0$ remains, and is of special significance.

To understand the level crossing consider a low energy theory for $E\ll\Delta_0$.
The two
eigenvectors $\xi_{\pm}$ of (\ref{h0}-\ref{dm}) with energy $\pm \epsilon_0(\phi)$ define Bogoliubov
operators $\Gamma_{0\pm} = \Psi^T \xi_{0\pm}$.
Due to particle-hole symmetry,
$\Gamma_{0+} = \Gamma_{0-}^\dagger \equiv \Gamma_0$.
The low energy Hamiltonian is thus
\begin{equation}
H =  \epsilon_0(\phi) (\Gamma_{0}^\dagger \Gamma_{0} - 1/2) = 2 i
\epsilon_0(\phi) \gamma_1 \gamma_2,
\label{hgamma}
\end{equation}
where we introduced Majorana operators $\gamma_1 =
(\Gamma+\Gamma^\dagger)/2$, $\gamma_2 = -i(\Gamma-\Gamma^\dagger)/2$.  In
the weak tunneling ($D\ll 1$) limit $\gamma_{1,2}$ describe the
Majorana end states at $x=\pm L/2$, which are weakly coupled by tunneling
of electrons.  Eq. \ref{hgamma}
describes two states distinguished by $N_0 \equiv \Gamma_0^\dagger \Gamma_0= 0,1$.
Mixing these states requires an interaction that changes $N_0$.
Due to the pairing term in (\ref{h0}), the total charge is not conserved.
However, the fermion parity, defined as the
number of electrons modulo 2, {\it is} conserved in (\ref{h0}-\ref{dm}).  This
forbids the coupling between the two states and protects the
crossing at $\epsilon_0(\phi)=0$.

There is a problem, however, with the fermion
parity. The junction Hamiltonian  (\ref{h0}-\ref{dm}) is invariant under a $2\pi$ phase
change,
but when $\phi\rightarrow \phi+2\pi$, the system passes through a single level crossing
and can only return to the initial state by a process which
changes $N_0$ by $1$.  The fermion parity thus
apparently changes when $\phi\rightarrow \phi+2\pi$.
This
has to do with the unbounded spectrum as $E\rightarrow -\infty$ and reflects a
{\it fermion parity anomaly} similar to the
$SU(2)$ anomaly in 4D field theory\cite{witten}.  This anomaly is
related to non-Abelian statistics.  When $\phi$
advances by $2\pi$  $\gamma_1
\rightarrow \gamma_1$ and $\gamma_2 \rightarrow -\gamma_2$.  In the
tunneling limit this can be interpreted as Ivanov's rule\cite{ivanov} for
the effect of braiding a vortex between the Majorana bound states.

The physical origin of the fermion parity anomaly lies in the topological structure of the
QSHI.  Consider first the Corbino disk in Fig. 1 without the
superconductor.  In Ref. \onlinecite{fukane1} we showed that the $Z_2$
invariant characterizing the QSHI describes the change in the $Z_2$
``time reversal polarization" (TRP) when flux $h/2e$ is threaded
through the hole.  A nonzero TRP specifies
a many body Kramers degeneracy
localized at either edge of the disk.  Since an odd number of fermions has a
Kramers degeneracy, the TRP is precisely the fermion parity.  With
the superconductor present, start in the groundstate at $\Phi=0$.
When flux $\Phi = h/2e$ is threaded through the hole,
$\phi$ advances by $2\pi$ {\it and} a unit of fermion parity is transferred
from the inner edge of the disk to the junction on the outer edge.
%
%
The anomaly in (1-3) is similar to the chiral anomaly for edge states
in the quantum Hall effect, where bulk currents
violate charge conservation at the edge.  Note however that though
(1-3) is
invariant under $\phi \rightarrow \phi+2\pi$, the
global Hamiltonian, which includes the bulk QSHI, is physically
distinct when $\Phi=0$ and $h/2e$.

The local conservation of fermion parity has important consequences
for the current and noise in a S-QSHI-S junction.  This
 is most striking near the degeneracy point for $\epsilon_0 \ll
\Delta_0$ and $T \ll \Delta_0$.  For the remainder of the paper we will
focus on that regime.  We will also consider the limit $L \ll v/\Delta_0$,
where there is a single Andreev bound state and (\ref{epsilon0}) applies,
though the results can straightforwardly be
generalized to the case with multiple Andreev levels, provided $T
\ll v/L$.    In this case, $N_0$
distinguishes two states, with Josephson currents
$I_\pm = \pm I_0$ with
\begin{equation}
I_0(\phi) =  {1\over 2}\sqrt{D} e \Delta_0 \sin \phi/2.
\label{i0}
\end{equation}
In the absence of transitions that violate local  fermion parity conservation
there can be no transitions between $I_+$ and
$I_-$, signaling a fractional Josephson effect.

Elastic scattering processes can be incorporated into the BdG
Hamiltonian from the start, and will not lead to violations of the
fermion parity.
However, at finite temperature, inelastic processes\cite{averin2,averin3} can lead to a transition
between $I_+$ and $I_-$, provided an
available fermion is present to switch the fermion parity.  This
could be either due
to a thermally excited quasiparticle or due to hopping from a bulk localized
state.  These processes, however, will be exponentially suppressed
at low temperature.  On a time scale longer than the switching
time the current will thermalize, with an average value\cite{ko,kwon}
\begin{equation}
\langle I(\phi) \rangle = I_0(\phi) \tanh \epsilon_0(\phi)/2T.
\label{iphi}
\end{equation}
On shorter times, the current will exhibit telegraph noise, as
it switches between $I_\pm$.

In order to model the inelastic processes responsible for the
telegraph noise we consider the interaction of the Andreev level
$\Gamma_0$ with a bath of fermions $c_n$ (e.g. quasiparticles) and bosons $b_m$ (e.g.
phonons).  We thus write
\begin{eqnarray}
H = \epsilon_0 \Gamma_0^\dagger \Gamma_0 + \sum_n E_n c_n^\dagger c_n
+ \sum_m \omega_m b_m^\dagger b_m \nonumber \\
 +\sum_{mn} \left[(V^1_{nm} c_n^\dagger b_m  + V^2_{nm} c_n b_m^\dagger)
\Gamma_0 + h.c.\right].
\end{eqnarray}
Here $E_n,\omega_n>0$, and we have ignored terms which create (or
anihilate) both fermions and bosons.  The transition rates $\tau^{-1}_\pm(\epsilon_0,T)$ between
the states $N_0$ and $N_0\pm 1$ follow from Fermi's golden rule.  For
$\epsilon_0,T \ll \Delta_0$ we find
\begin{equation}
\tau^{-1}_\pm  = e^{\mp \epsilon_0/2T}\left(w_1(T)
e^{\epsilon_0/2T} + w_2(T) e^{-\epsilon_0/2T}\right),
\end{equation}
where
\begin{equation}
w_{1,2}(T) = 2\pi\sum_{n,m} e^{-E_n/T} |V^{1,2}_{nm}|^2
\delta(E_n-\omega_m).
\end{equation}
If either the Zeeman term vanishes ($M_0=0$) or the system is
symmetric under $x\rightarrow -x$, then $w_1(T) = w_2(T) \equiv
w(T)$.  We will assume this below, though the results are only
slightly modified otherwise.  $w(T)$
depends on the dominant source of fermions, which we take to be either thermally
activated quasiparticles or Mott variable range hopping from bulk localized states.
\begin{equation}
w(T) \propto \left\{\begin{array}{ll}
e^{-\Delta_0/T} & {\rm quasiparticles}, \\
e^{-(T_0/T)^{1/3}} & {\rm hopping.}
\end{array}\right.
\label{rho}
\end{equation}
$T_0$ depends on the density of states and
localization length, and we assume the hopping is 2D.

The transition rate is exponentially suppressed for $T\rightarrow 0$.
At sufficiently low temperature the
resulting telegraph noise could be observed in the time domain.  At higher
temperature there is a signature in the noise
spectrum $S(\omega)$.  We determine $S(\omega)$ semiclassically by solving
a kinetic equation for the probability $p(t)$ that
$N_0=1$\cite{averin2,averin3}.
This has the form
$dp/dt = - (p - \bar p)/\tau$, where $\tau^{-1} = \tau_+^{-1} + \tau^{-1}_- = 4 w
\cosh^2 \epsilon_0/2T$.  $\bar p = (1+\exp\epsilon_0/T)^{-1}$ follows from the detailed balance
condition $\tau_+/\tau_- = e^{\epsilon_0/T}$.  Temporal correlations in $I(t)$ decay
exponentially on a time scale $w^{-1}$, and the noise spectrum
$S(\omega) = 2 \int_{-\infty}^\infty e^{i\omega t}\langle I(t) I(0)\rangle$
is given by\cite{averin1,averin2}
\begin{equation}
S(\omega) =
{4 I_0^2 \over {\cosh^2 \epsilon_0(\phi)/2T}}{\tau\over{1+\omega^2\tau^2}}.
\label{sw}
\end{equation}
In the zero frequency limit we have
\begin{equation}
S(\omega\rightarrow 0 ) = {I_0^2\over{w(T) \cosh^4
\epsilon_0(\phi)/2T}}.
\label{s0}
\end{equation}

For $D=1$, these results are similar to the theory of a
SQPC\cite{ko,averin2,averin3,averin1}.
However, the current in Eq. (\ref{i0}) is
{\it half} the value of a perfect single channel SQPC.   A SQPC is like two copies of a
S-QSHI-S junction.  This leads to a fundamental difference, because in the SQPC
there is no conservation law to prevent scattering between the
$\pm I_0$ states, which can occur via low energy processes that transfer an electron
between the two pairs.  Elastic backscattering in the SQPC
leads to an avoided crossing of
the states near $E=0$, and the Andreev states carry no current at $\phi=\pi$.  Even if the transmission of the SQPC
 is perfect, inelastic processes will
couple the states.  For instance, near the degeneracy point $\epsilon_0(\phi)=0$
spin flip scattering via the nuclear hyperfine interaction will lead
to a finite lifetime for transitions, even at low temperature.
It is also of interest to compare with the theory
of Ref. \onlinecite{kwon}.  In that work, multi channel junctions were
considered.   Independence of the different channels requires translational
symmetry, so impurity scattering will lead to the violation of
the conservation fermion parity within a given channel.  Thus the low temperature behavior
predicted by (\ref{rho}) and (\ref{s0}) is unique to the S-QSHI-S junction, and is a signature of the
fermion parity anomaly.

We now briefly consider junctions at finite voltage
bias.  The analysis is similar to Ref. \onlinecite{averin1}.
There are two cases to
consider, depending on $M_0$.  For $M_0=0$, the perfect transmission of the edge states causes
the Andreev levels to merge with the continuum levels.  This leads to
a finite DC current, which for $eV \ll \Delta_0$
can be understood semiclassicallly in terms of multiple Andreev reflections.  For
$w(T) \ll eV \ll \Delta_0$, the current is $I(V) = (2/\pi)I_c {\rm sgn}V$, where
$I_c = \sqrt{D} e\Delta_0/2$.
For $M_0>0$, there is an energy gap $\delta$ separating the Andreev levels
from the continuum, as in Fig. 2(b,d).  For $w(T) \ll eV \ll \delta$ there will be a fractional AC
Josephson current with frequency $eV/\hbar$\cite{kwon}.   For $eV \sim \delta$ Landau-Zener
tunneling
processes through the gap $\delta$ will lead to a damping of the AC Josephson current as well as a
finite DC current.

We close by discussing the feasibility of experiments using the QSHI recently achieved in
HgCdTe quantum wells\cite{bhz,konig}, which
has a bulk gap of order 20 meV\cite{zreview}.  The desired geometry would be similar to Ref. \cite{inas},
where a 2D InAs quantum well was contacted with Nb. The proximity induced
gap will depend on the contact.
If optimized $\Delta_0$ could be of order the bulk gap of the superconductor.
To determine the required junction size
we use $v=3.6$ eV\AA\cite{zreview} and $\Delta_0 = .1$meV.
Then, $L \lesssim  v/\Delta_0 \sim 3 \mu$m  sets the scale for
having a single Andreev level.
The simplest experiment would be to study a single current biased junction,
which is predicted to have a critical current $I_c = e\Delta_0/2 \sim 10 {\rm nA}$,
which is half the value of a perfect single channel SQPC.
Measuring the equilibrium noise at $\phi\sim \pi$ requires an
inductive measurement on a ring\cite{review2}.
The physics at $M_0 \ne 0$ requires a magnetic field
in the junction region.  For an
appropriately aligned field, a field induces a gap
$B \times (3.1{\rm meV/T})$\cite{zreview} in the edge states, so a field of order
$.03 T$ could suppress the normal state transmition $D$ as well as the magnitude of the
Josephson current.

It is a pleasure to thank Patrick Lee and Erhai Zhao for
helpful discussions.   This work was supported by NSF grant
DMR-0605066 and ACS PRF grant 44776-AC10.

\end{document}